\documentclass[preprint,12pt]{elsarticle}

\usepackage{graphicx}
\usepackage{amssymb}
\usepackage{natbib}
\usepackage{units}
\usepackage{bm}
\usepackage{dcolumn}

\usepackage{lineno}

\def\Neff{N_{\rm eff}}
\def\meff{m_{\rm eff}^{\rm sterile}}
\def\DNeff{\Delta N_{\rm eff}}
\journal{Elsevier}

\begin{document}

\begin{frontmatter}


\title{A Combined View of Sterile-Neutrino Constraints from CMB and Neutrino Oscillation Measurements}



\author{Sarah~Bridle}
\author{Jack~Elvin-Poole}
\author{Justin~Evans}
\author{Susana~Fernandez}
\author{Pawel~Guzowski}
\author{Stefan~S\"oldner-Rembold}
\address{The University of Manchester, School of Physics and Astronomy, Manchester, M13 9PL, United Kingdom
}

\begin{abstract}
We perform a comparative analysis of constraints on sterile neutrinos from the Planck experiment and from current and future neutrino oscillation experiments (MINOS, IceCube, SBN). For the first time, we express joint constraints on $\Neff$ and $\meff$ from the CMB in the $\Delta m^{2}$, $\sin^{2} 2 \theta$ parameter space used by oscillation experiments. We also show constraints from oscillation experiments in the $\Neff$, $\meff$ cosmology parameter space. In a model with a single sterile neutrino species and using standard assumptions, we find that the Planck 2015 data and the oscillation experiments measuring muon-neutrino
($\nu_{\mu}$) disappearance have similar sensitivity. 

\end{abstract}


\end{frontmatter}


\section{Introduction}
\label{intro}

The search for low-mass sterile neutrinos is motivated by several experimental anomalies that are not consistent with the three-flavour paradigm. Sterile neutrinos would change the oscillation probabilities observed by detecting neutrinos from accelerators, nuclear reactors, or produced in the atmosphere. On a cosmological scale, they would modify the power spectrum of the Cosmic Microwave Background (CMB)
(Fig. \ref{fig:cmb_oscillations}). 

Both types of measurement put severe constraints on the existence of extra neutrino flavours, but they are evaluated in terms of different parameter sets. The CMB measurements constrain the effective number of additional neutrino species, $\DNeff$ (above the Standard Model (SM) prediction of $N_{\rm eff} = 3.046$), and the effective sterile neutrino mass $\meff$. Oscillation experiments parameterize their constraints in terms of mass-squared differences, $\Delta m^{2}_{ij}$, between the mass eigenstates, and the mixing angles $\theta_{\alpha\beta}$ between mass and flavour eigenstates. Here, we use the calculation of~\cite{Hannestad2012} and show the Planck CMB cosmology constraints in the same parameter space as used for $\nu_{\mu}$ disappearance measurements.

Several experimental anomalies related to the appearance and disappearance of $\nu_e$ could be explained by light sterile neutrinos with a mass-squared difference relative to the active states of $\Delta m^{2} \approx \unit[1]{eV^{2}}$~\cite{Abazajian2012,Collin:2016rao,Collin2016_globalfit}. The LSND Collaboration observes an excess of $\bar{\nu}_{e}$ appearance in a $\bar{\nu}_{\mu}$ beam~\cite{LSND}, and MiniBooNE measures an excess of both $\nu_e$~\cite{Miniboone2007} and $\bar{\nu}_{e}$ appearance~\cite{Miniboone2010,Miniboone2013}. Reactor experiments observe a deficit of $\approx 6\%$ in the $\bar{\nu}_{e}$ flux compared to expectations~\cite{reactor}. Furthermore, Gallium experiments observe a smaller 
$\nu_e+\mbox{$^{71}$Ga}\to \mbox{$^{71}$Ge}+e^{-}$ event rate than expected from
$^{51}$Cr and $^{37}$Ar sources~\cite{Giunti:2010zu}.
The Daya Bay Reactor experiment has searched for $\bar{\nu}_e$ disappearance setting limits on the mixing angle $\sin^2\theta_{14}$ in the low
$\Delta m^{2}$ region $0.0002<\Delta m_{41}^{2}< \unit[0.2]{eV^{2}}$~\cite{An:2016luf}. These results have been combined with $\nu_{\mu}$ disappearances searches by MINOS~\cite{MINOS:2016viw} to obtain stringent constraints 
on the product $\sin^22\theta_{14}\sin^2\theta_{24}$~\cite{Adamson:2016jku}. 
For this analysis, we focus on recent $\nu_{\mu}$ disappearance results, where no anomalies have been found, and assume that $\sin^2\theta_{14}= \sin^2\theta_{34}=0$ in order to be consistent with the assumptions that were used for deriving these limits.

Several studies have combined oscillation and cosmological data to constrain sterile neutrinos. Several~\cite{ref:Archidiacono2012,ref:Archidiacono2013,Gariazzo2013,Archidiacono2014,Archidiacono2016} use the posterior probability distribution on $\Delta m^{2}$ from short-baseline anomalies as a prior in the cosmological analysis. Here, we convert the full CMB cosmology constraints into the oscillation parameterisation and vise versa, focusing on recent $\nu_{\mu}$ disappearance results. This conversion has also been studied in~\cite{Mirizzi2013,Melchiorri2009}. 
Our analysis differs in several ways: (i) unlike~\cite{Mirizzi2013} we use the 2D combined constraints on $\DNeff$ and $\meff$ in the cosmological analysis, rather than converting 1D constraint values in each parameter individually; (ii) we use the latest CMB data from Planck, updating from the WMAP 5-year data used in~\cite{Melchiorri2009}; (iii) we solve the full quantum kinetic equations, rather than using the averaged momentum approximation~\cite{Mirizzi2012} used in~\cite{Mirizzi2013,Melchiorri2009}; (iv) we also consider the impact of non-zero lepton asymmetry, $L$, and a different sterile mass mechanism. The lepton asymmetry is defined as ${L=(n_{f}-n_{\bar f})N_{f}/N_{\gamma}}$, where $n_{f}$ and $n_{\bar f}$ are the number densities of fermions and anti-fermions, respectively, and $N_{f}$ and $N_{\gamma}$ are the numbers of fermions and photons.

\section{Data sets}
\label{data sets}

\subsection{Cosmological Data sets}

Observations of the CMB radiation are the most powerful probe of cosmology, giving a snapshot of the Universe around 300,000 years after the Big Bang.
The angular intensity fluctuations are sourced by temperature fluctuations in the plasma, which in turn depend on the constituents of the Universe, including sterile neutrinos. 
Cosmology results are most sensitive to the sum of all neutrino masses, rather than the relative masses of the active and sterile neutrinos.
The Planck Satellite currently provides the definitive measurement of the CMB temperature anisotropies~\cite{planck_overview}.
The Planck data have been used to constrain the sum of the active neutrino masses yielding $\sum m_{\nu} < \unit[0.68]{eV}$ from CMB temperature data alone \cite{Planck_cosmo}.
The information from the CMB can also be combined with that from other cosmological observations for even tighter constraints~\cite{Dvorkin, MacCrann, Mantz, Battye2014, Huang2015, Cuesta2015, Palanque-Delabrouille2015, Valentino2015}. Here, we use the Planck temperature power spectrum and low multipole polarisation data alone.

\begin{figure*}
  \centerline{
  \scalebox{1.}{
  \includegraphics[scale=0.37]{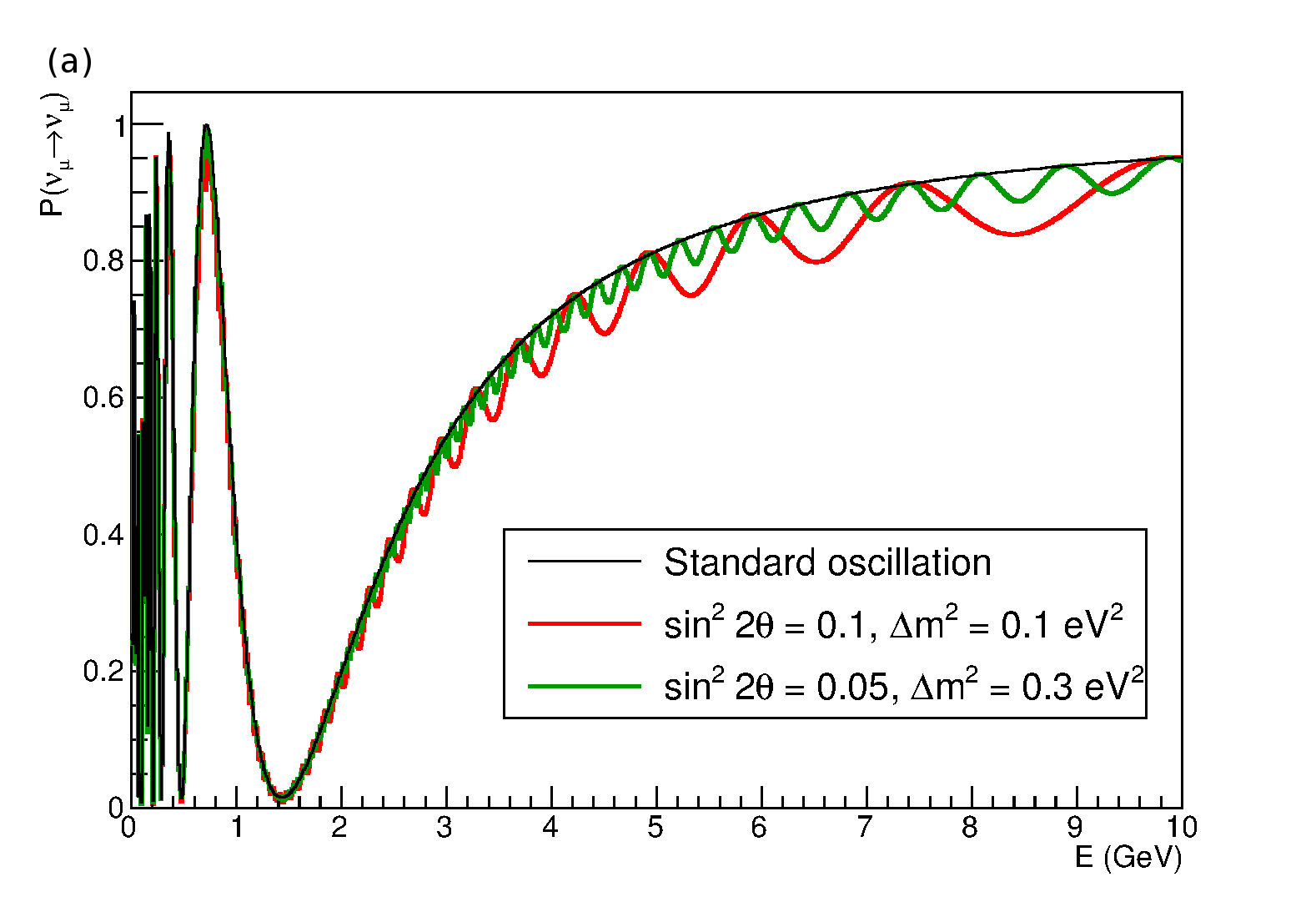}
  }
  \scalebox{1.}{
  \includegraphics[scale=0.31]{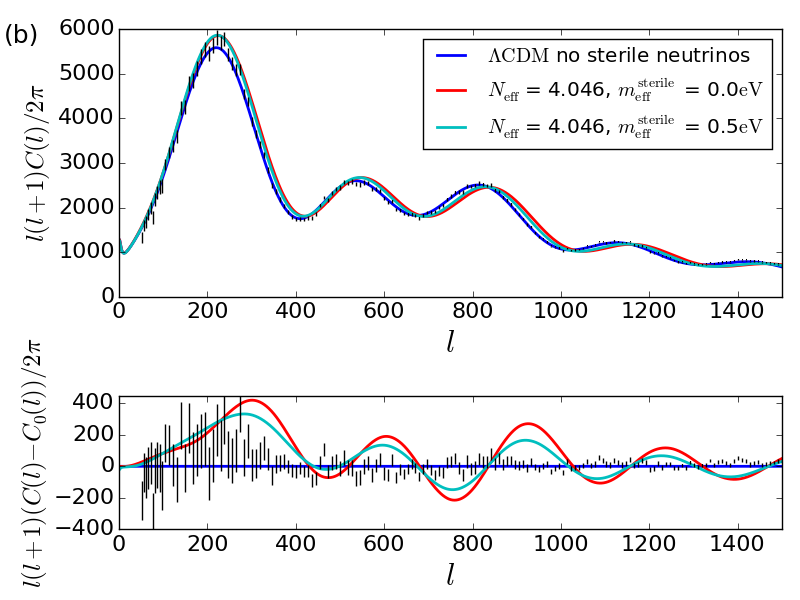}
  }
  }
  \caption{
  (a) The $\nu_{\mu}$ survival probability after \unit[735]{km} as a function of neutrino energy, for different oscillation parameter values.
  (b) The CMB temperature power spectrum for different values of the effective number of neutrino species and effective sterile mass, fixing the cold dark matter energy density. The $\Lambda {\rm CDM}$ case uses the SM value $N_{\rm eff} = 3.046$ and an active neutrino mass sum $\sum m_{\nu} = \unit[0.06]{eV}$. All other cosmological parameters are set to the best-fit values from the Planck 2015 data shown by error bars \cite{planck_cls}. The power spectra are generated using the CAMB module~\cite{camb} in CosmoSIS~\cite{cosmosis}.}
  \label{fig:cmb_oscillations}
\end{figure*}

To constrain sterile neutrinos, two parameters are added to the baseline Planck analysis: the effective sterile mass, $\meff = \unit[(94.1 \: \Omega_{\rm sterile} h^2)]{eV}$, and the effective number of additional neutrino species, $\DNeff = \Neff - 3.046$. The cosmological model used is $\Lambda {\rm CDM} + \meff + \DNeff$. Additional cosmological parameters and their degeneracies with neutrino parameters are not considered here.

Figure~\ref{fig:cmb_oscillations} (b) shows the power spectrum of the CMB temperature fluctuations. We observe that increasing the effective number of neutrino species, while fixing $\meff = 0$, shifts the peak structure to higher multipoles, $l$, due to a change in the matter-radiation equality redshift, $z_{\rm eq}$. There is also
an increase in the integrated Sachs-Wolfe effect at low $l$~\cite{Archidiacono2013}. 
A non-zero $\meff$ further changes $z_{\rm eq}$ adding to the shift of the peak locations~\cite{Gariazzo2015,Abazajian2012,Archidiacono2013,Lesgourgues2014}.

The effective mass, $\meff$, can be related to the mass of the sterile neutrino, $m_{\rm sterile}=m_4$,
in two ways. The first is to assume a thermal distribution with an arbitrary temperature $T_{s}$. The quantity $\DNeff$ is then a measure of the thermalisation of the sterile neutrinos, $\DNeff = (T_{s}/T_{\nu})^4$, yielding,
\begin{equation}
\meff = \left(\frac{T_{s}}{T_{\nu}}\right)^{3} m_{\rm 4}^{\rm thermal} = (\DNeff)^{3/4} m_{\rm 4}^{\rm thermal} \, . \\
\label{eqn:thermal}
\end{equation}
The second model assumes the extra eigenstate is distributed proportionally to the active state by a scaling factor, $\chi_{s}$, here equal to $\DNeff$,
\begin{equation}
\meff = \chi_{s} m_{\rm 4}^{\rm DW} = \DNeff m_{\rm 4}^{\rm DW} \, .
\label{eqn:dw}
\end{equation}
This is known as the Dodelson-Widrow (DW) mechanism~\cite{Dodelson1993}. 
We use the thermal distribution as our fiducial interpretation and show that our conclusions are robust to this choice. 

\begin{figure*}
  \centerline{
  \scalebox{1.}{
  \includegraphics[scale=0.38]{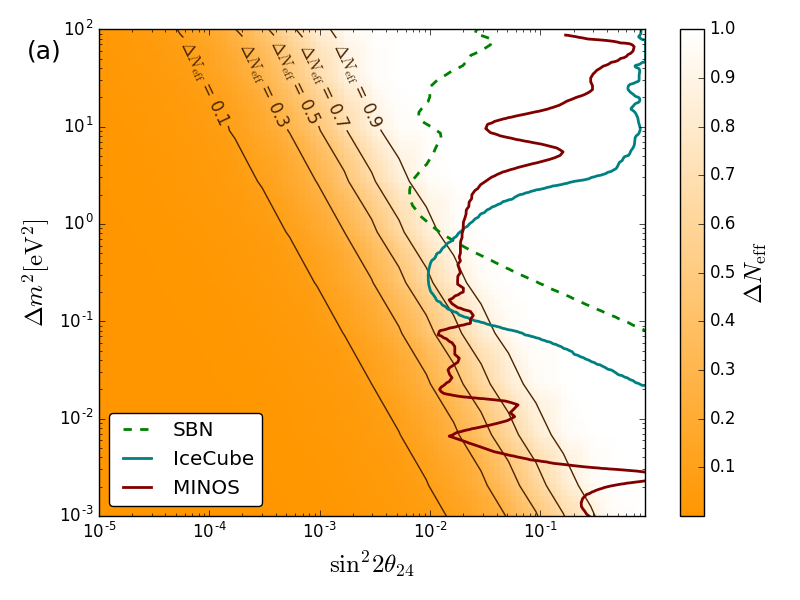}}
  \scalebox{1.}{
  \includegraphics[scale=0.38]{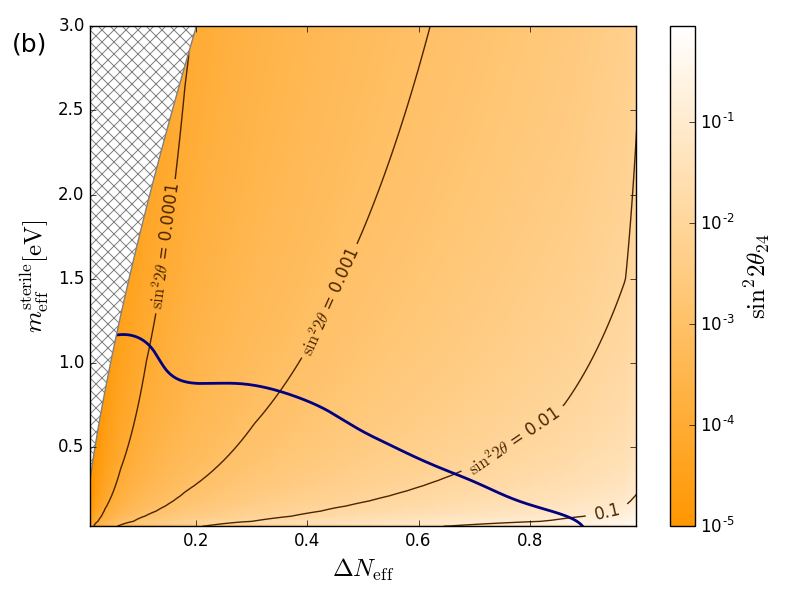}}
  }
  \centerline{
  \scalebox{1.}{
  \includegraphics[scale=0.38]{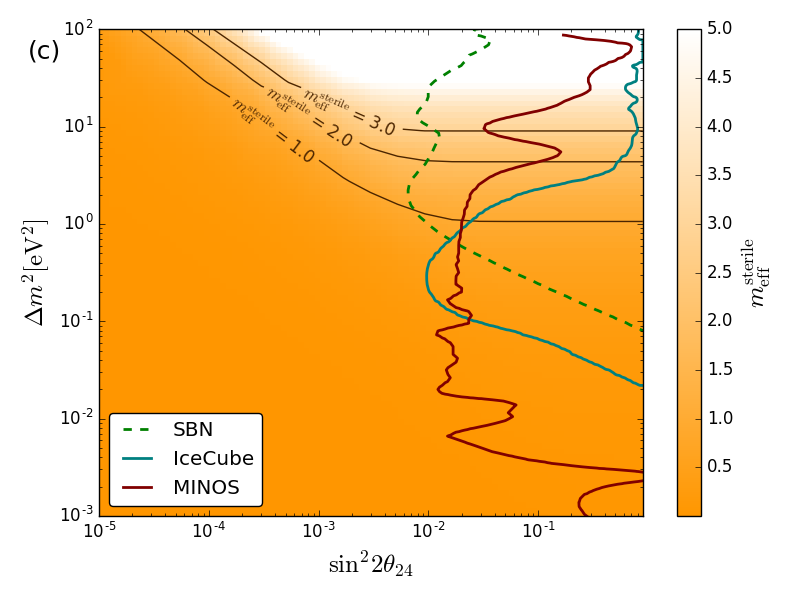}}
  \scalebox{1.}{
  \includegraphics[scale=0.38]{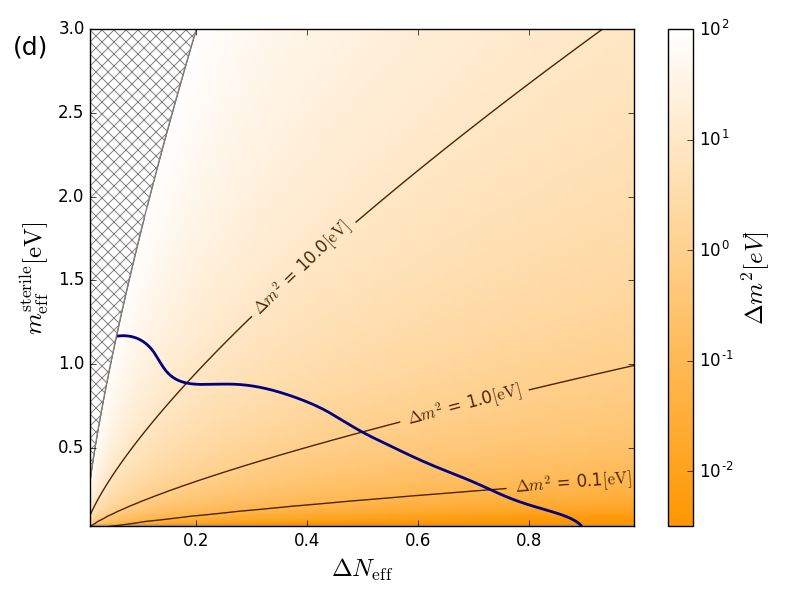}}
  }
  \caption{(a, c) Cosmological parameters $\Delta N_{\rm eff}$ and $\meff$ calculated in the oscillation space $\Delta m^{2}$, $\sin^{2} 2 \theta_{24}$ using LASAGNA. We use the thermal sterile neutrino mass (Eq.~\protect\ref{eqn:thermal}) and $L=0$. Also shown are the constraints from the experiments native to this space, MINOS and IceCube, and the SBN sensitivity. The region to the right of the contours is ruled out at the $95\%$ CL.
  (b, d) $\Delta m^{2}$, $\sin^{2} 2 \theta$ in the cosmological space, $\meff$, $\DNeff$. The region above the blue line is excluded by the Planck temperature and low-$l$ polarization data at $95\%$ CL. A prior of $m_{4}^{\rm thermal} < \unit[10]{eV}$ has been applied as in~\cite{Planck_cosmo}. The hatched area corresponds to $m_{4}^{\rm thermal} > \unit[10]{eV}$ where $\DNeff$ was not calculated.}
  \label{interp_matrix}
\end{figure*}

The Planck analysis assumes the normal mass ordering of the active neutrinos with the minimum masses allowed by oscillation experiments, $m_{1} = \unit[0]{eV}$, $m_{2} \approx \unit[0]{eV}$, and $m_{3} = \unit[0.06]{eV}$. Any excess mass is considered to be from a single additional state, which
implies that $\Delta m_{41} ^2 \approx m_{4}^2$. We use these assumptions throughout our analysis. Assuming inverted mass ordering or allowing $m_{1} > 0$ would strengthen the Planck constraints on sterile neutrinos. These assumptions allow us to directly compare to the oscillation data. 
The Planck $95\%$ Confidence Level (CL) contour is shown in Fig.~\ref{interp_matrix} (b, d) for a prior of $m_{4}^{\rm thermal} < \unit[10]{~eV}$~\cite{Planck_cosmo}. 

\subsection{Oscillation Data sets}

The MINOS experiment~\cite{minos1} reconstructs interactions from a $\nu_{\mu}$ beam created in an accelerator at Fermilab in a near detector (ND), located about \unit[1]{km} from the source, and a far detector (FD) at \unit[735]{km}. A sterile neutrino will reduce the $\nu_{\mu}$ survival probability through its mixing with the active
neutrinos (Fig.~\ref{fig:cmb_oscillations} (a)). In most analyses, the ND serves as a reference point that defines the un-oscillated beam
spectrum. However, for mass differences above $\Delta m^2\approx \unit[1]{eV^{2}}$, oscillations occur rapidly and can already lead to a depletion of the neutrino flux at the ND. MINOS has therefore performed an innovative analysis exploiting the ratio of the neutrino energy spectra measured in the FD to those in the ND using both charged-current (CC) $\nu_{\mu}$ and neutral-current (NC) neutrino interactions~\cite{MINOS,MINOS:2016viw}.
Limits on sterile-neutrino parameters are obtained by performing a $\chi^2$ fit of the far-over-near ratio for both CC and NC data samples. 

We use the $\chi^2$ surface given in~\cite{MINOS}, which includes the data published in~\cite{MINOS:2016viw} and incorporates the statistical
uncertainties, a full covariance matrix of the experimental systematic uncertainties, and a weak constraint on $\Delta m_{32}^2$, which the data can then itself constrain. All other three-flavour oscillation parameters are fixed in the MINOS fit. We assume that all uncertainties follow a Gaussian distribution, and derive confidence levels using Gaussian $\chi^{2}$ $p$-values. The $95\%$ CL contour derived from the MINOS $\chi^2$ distribution is shown in Fig.~\ref{interp_matrix} (a, c).

The IceCube detector~\cite{icecube1} comprises 5160 optical modules instrumenting $\sim\!\! \unit[1]{km^3}$ of ice at the South Pole. 
Neutrinos are detected using Cherenkov radiation emitted by charged particles produced in CC interactions. 
This is used to measure the disappearance of atmospheric muon neutrinos
($\nu_{\mu}$ and $\bar{\nu}_{\mu}$) that have traversed the Earth. Sterile neutrinos are expected to modify the energy-dependent zenith-angle distribution of the $\nu_{\mu}$ and $\bar{\nu}_{\mu}$ through resonant matter-enhanced oscillations caused by the MSW effect~\cite{ref:Wolfenstein,ref:Mikheev}. IceCube has searched for sterile neutrinos by studying the 2D distribution of the reconstructed neutrino energy and zenith angle~\cite{TheIceCube:2016oqi,BenJones}. 

The IceCube likelihood distribution utilizes both shape and rate information, including systematic and statistical uncertainties. The distributions shown
in Fig.~\ref{interp_matrix} (a, c) are taken from~\cite{BenJones}. 
The IceCube Collaboration also assumes $\theta_{34}=0$ in its analysis. It shows that this assumption leads to a more
conservative limit and that non-zero values of $\theta_{14}$ have
little effect on the results~\cite{TheIceCube:2016oqi}.

The Short Baseline Neutrino (SBN) programme~\cite{SBNproposal} at Fermilab will study the LSND~\cite{LSND} and MiniBooNE~\cite{Miniboone2007,Miniboone2010,Miniboone2013} anomalies. It comprises three liquid-argon time projection chambers at different baselines in a $\nu_{\mu}$ beam: the already-running MicroBooNE detector, and the SBND and ICARUS detectors that are due to start data-taking in 2018. The programme will primarily search for $\nu_e$ appearance, but can also study the disappearance of $\nu_\mu$.

We use GLOBES~\cite{globes} to estimate the SBN sensitivity in the $\nu_\mu$ disappearance channel. The GENIE Monte Carlo (MC) generator~\cite{GENIE} is used to calculate the $\nu_\mu$ CC interaction cross section on argon. We develop a toy MC model to calculate the geometric acceptance, using the GENIE output of muon momentum and direction, muon range tables~\cite{MuonRange}, and interaction vertices distributed uniformly inside the active dimensions of the detectors. Acceptances are calculated for fully-contained muons, or for exiting muons with a track length of at least \unit[$1$]{m} inside the detector. Energies of contained muons are smeared by a $2\%$ absolute resolution, while for exiting muons the resolution is assumed to be $10\%/\sqrt{E [\textrm{GeV}]}$. Hadronic energy is smeared by $20\%/\sqrt{E [\textrm{GeV}]}$. We apply an overall selection efficiency of $80\%$.

We combine the beam fluxes from Fig.~3 of~\cite{SBNproposal} with the modeled efficiencies and energy-smearing matrices to provide inputs to GLOBES, which is used to calculate the $\chi^2$ surface as a function of the mixing angle $\theta_{24}$ and mass-squared difference $\Delta m^2_{41}$. We always set $\theta_{14}=\theta_{34}=0$.
Only the $\nu_{\mu}$ disappearance channel is used to make a direct comparison with the MINOS and IceCube measurements in Figs.~\ref{interp_matrix} (a, c).

\section{Thermalisation of sterile neutrinos}

To relate the cosmological parameterization \mbox{($\meff$, $\DNeff$)} to the oscillation parameterization \mbox{($\Delta m_{ij} ^2$, $\theta_{\alpha \beta}$)}, we solve the full quantum kinetic equations that govern the sterile neutrino thermalization~\cite{Hannestad2012}. We use LASAGNA~\cite{Hannestad2013} to solve these equations in the simplified scenario with one active and one sterile neutrino flavour as described in~\cite{Enqvist1992,Stodolsky1987}. This scenario contains a single mixing angle, $\theta$, and the flavour states are
\begin{eqnarray}
\nu_a = \cos \theta \nu_1 - \sin\theta \nu_2 \, ,\\
\nu_s = \sin \theta \nu_1 + \cos\theta \nu_2  \, ,
\end{eqnarray}
where $\nu_{1,2}$  are the mass eigenstates, and $\nu_{a,s}$ the active and sterile flavour eigenstates, respectively. 

The LASAGNA input parameters are the mass splitting, $\Delta m^2$, between the two mass states, the mixing angle, $\theta$, the lepton asymmetry, $L$, and the range in temperature, $T$, over which to evolve the kinetic equations.
LASAGNA produces a grid in the parameter $x = p/T$, where $p$ is the neutrino momentum, upon which the factor 
\begin{equation}
P_{s}^{+} = (P_{0} + \bar{P}_{0}) + (P_{z} + \bar{P}_{z})  
\end{equation}
is calculated. Here,
$P_{0}$ and $P_{z}$ are the first and last components of the neutrino Bloch vector, $(P_{0},P_{x},P_{y},P_{z})$. The factor $P_{0}$ corresponds to the number density of the mixed state, and $P_{z}$ is related to the probability that a neutrino is in the sterile or active state, \mbox{${\rm Prob}(\nu_s) = (1-P_{z})/2$}, and \mbox{${\rm Prob}(\nu_a) = (1+P_{z})/2$}. The factors $\bar P_{0}$ and $\bar P_{z}$ are the corresponding anti-neutrino values.
We use $P_{s}^{+}$ to calculate 
\begin{equation}
\DNeff = \frac{\int dx\, x^{3} f_{0} P_{s}^{+}}{4 \int dx\, x^{3} f_{0}}
\end{equation}
with the Fermi-Dirac distribution function, \mbox{$f_{0} = 1/(1+e^{x})$}. This is valid if the active states are in thermal equilibrium. More details on LASAGNA are given in~\cite{Hannestad2012,Hannestad2013} and on the quantum kinetic equations in~\cite{Enqvist1992,Stodolsky1987}. 

For our fiducial analysis, we run LASAGNA with $L=0$ in a temperature range $1<T<\unit[40]{MeV}$, calculating $\DNeff$ on the 2D grid of $\Delta m^{2}$, $\sin^{2} 2 \theta$ values shown in Fig.~\ref{interp_matrix} (a). 
We convert positions in the cosmology parameter space ($m_{\rm eff}^{\rm sterile}$, $\Delta N_{\rm eff}$) into the oscillation space ($\Delta m_{41}^{2}$, ${\sin}^{2} 2 \theta_{24}$), first by using $\Delta m_{41} ^2 = m_{4}^2$ and Eq.~\ref{eqn:thermal}
 to find $\Delta m_{41}^{2}$, then interpolating $\sin^{2} 2 \theta_{24}$ from the underlying grid in Fig.~\ref{interp_matrix}. We assume that the sterile-active mixing is dominated by a single angle $\theta_{24}$.
 
\section{Results}
\label{sec:results}

\begin{figure*}
\centerline{
\scalebox{1.}{
\includegraphics[scale=0.53]{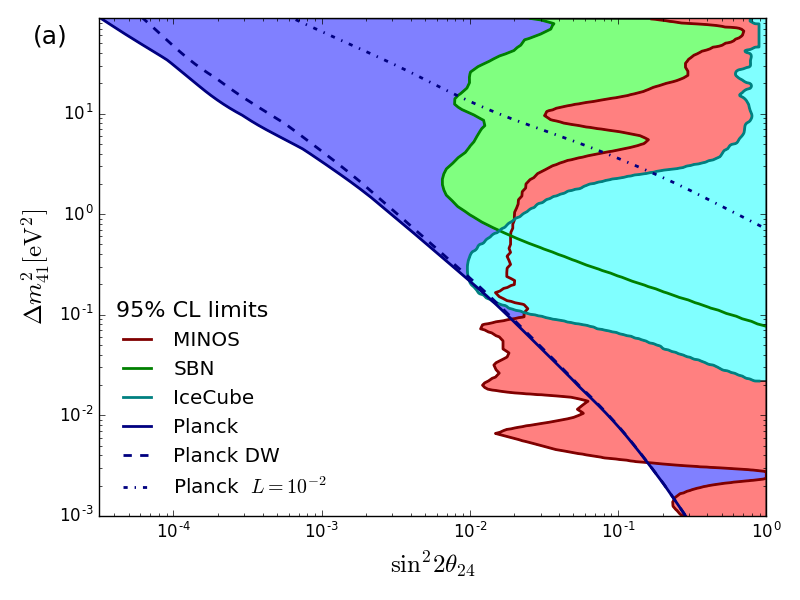}}}
\centerline{
\scalebox{1.}{
\includegraphics[scale=0.53]{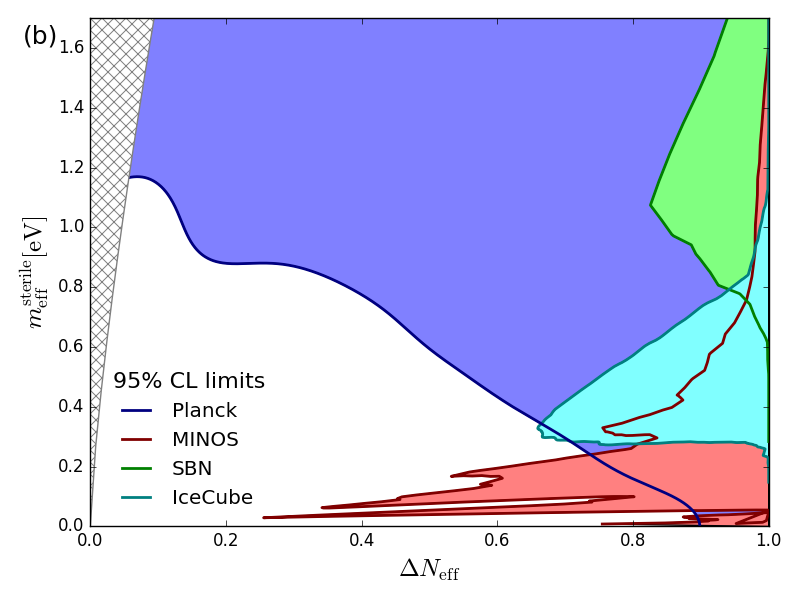}}}
\caption{(a) Sterile neutrino exclusion regions at $95\%$ CL from Planck, MINOS, IceCube, and the SBN forecast in the oscillation parameter space. The dashed line is the Planck constraint with $m_{4}$ calculated using the Dodelson-Widrow mechanism. The dot-dash line is the Planck constraint using a large lepton-asymmetry, $L=10^{-2}$. 
(b) The same contours in the cosmological space, where the difference between the thermal and Dodelson-Widrow scenarios is negligible.
}
\label{fig:exclusion_regions}
\end{figure*}

The experimental inputs have been derived using different statistical approaches. For the oscillation experiments a $\Delta\chi^2$ contour is calculated as the difference of the $\chi^2$ of the best-fit hypothesis for the data to the $\chi^2$ at each model point.
For the Planck data, a multi-dimensional Markov Chain MC is produced allowing cosmological, nuisance and neutrino parameters ($\DNeff$, $\meff$) to be varied. The number density of points in this chain is proportional to the likelihood ${\cal L}$. We draw this likelihood surface in the ($\DNeff$, $\meff$) plane and take the $\chi^2$ to be $-2\ln({\cal L})$. 
The contour describing the $95\%$ CL
corresponds to $\chi^2 - \chi_{\rm min}^2=5.99$ in the 2D input distributions. In the Planck case, this leads to a dependence on the prior of these parameters. In our analysis the priors are flat in the ranges $0 < \DNeff < 1$, and $0 < \meff < \unit[3]{eV}$.

We make several standard assumptions about the cosmological and neutrino models that may impact our conclusions. We assume $\Lambda_{\rm CDM}$ in the Planck analysis and a single sterile neutrino species, mixing only by one channel, in our conversion between parameter spaces. We also assume that any $\DNeff$ is caused only by neutrinos and no other light relic particle. Studying the impact of these assumptions is beyond the scope of this paper.

Fig.~\ref{fig:exclusion_regions} shows the CMB and oscillation experiment exclusion regions on the same axes in the oscillation and cosmology parameter spaces. The CMB data excludes a similar corner of the parameter space to the oscillation experiments, ruling out large mixing angles and large sterile-neutrino masses within the 3+1 model. 
This conclusion is unchanged by switching from the thermal mass in Eq.~(\ref{eqn:thermal}) to the Dodelson-Widrow mechanism in Eq.~(\ref{eqn:dw}). 

The condition $\Delta m_{41}^{2} < 10^{-2.4}$, leads to $m_{4}<m_{3}$, and the active masses can no longer be treated as a single state. Therefore, the Planck contour below this value is too conservative. Ref.~\cite{Mirizzi2013} discusses cosmological constraints in this $\Delta m_{41}^{2}$ range.

The Planck contour at large $\Delta m_{41}^{2}$ is dominated by the constraint on $\meff$, and at low $\Delta m_{41}^{2}$ by the constraint on $\DNeff$, as shown in Figs.~\ref{interp_matrix} (a, c). This is also illustrated in Fig.~3 of~\cite{Mirizzi2013}, which converts 1D upper limits on each of these parameters separately, instead of the 2D likelihood surface. 
Comparing these results to the averaged-momentum approximation results of~\cite{Mirizzi2013}, we find that solving the full quantum kinetic equations results in qualitatively similar constraints. 

In the fiducial analysis, we convert between parameter spaces using the assumption $L=0$. In this case, the Planck data is more constraining than the oscillation experiments for large mass-squared differences, $\Delta m_{41}^{2} > \unit[10^{-1}]{eV^{2}}$, and less constraining than MINOS in the range $10^{-2} < \Delta m_{41}^{2} < \unit[10^{-1}]{eV^{2}}$. When the lepton asymmetry is large, $L=10^{-2}$, the oscillations between sterile and active neutrinos are suppressed, giving a lower $\DNeff$ for the same oscillation parameters (see Fig.~4 of~\cite{Hannestad2012}). This weakens the Planck constraints in the oscillation space such that they are now less constraining than all of the oscillation experiments considered, as shown by the dot-dash line in Fig.~\ref{fig:exclusion_regions} (a).

The MINOS experiment is particularly sensitive to the region of low $\Delta m_{41}^{2}$ because of its baseline and neutrino energy range. This is the only region where the oscillation data are more constraining than the cosmology data when assuming $L=0$. In the cosmology space, this corresponds to ruling out a region of large $\DNeff$ at low $\meff$.  

\section{Conclusions}
\label{sec:conclusions}

In conclusion, we compare sterile neutrino constraints from oscillation experiments and cosmological constraints. We use the quantum kinetic equations to convert between the standard oscillation parameterization of neutrinos (the mass-squared difference and mixing angle) and the cosmology parameterization (the effective sterile neutrino mass and the effective number of neutrino species). We show the relationship between each of the parameter combinations. 

We show the Planck 2015 CMB cosmology constraints in the oscillation parameter space and find that they rule out large values of $\Delta m_{41}^{2}$ and mixing angle $\theta$. For the fiducial case, the region of parameter space ruled out by IceCube data is already excluded by the Planck CMB constraints. For the first time, we show that much of the MINOS exclusion region is also ruled out by Planck CMB constraints, although for low $\Delta m_{41}^{2}$ MINOS is more constraining. The forecast constraints for the SBN experiments are not expected to add to the information already provided by Planck CMB results with these model assumptions. However, their main sensitivity will be through the $\nu_e$ appearance searches not considered here. The MINOS data adds the most information to that provided by Planck CMB measurements because it probes the lowest $\Delta m^{2}$.

The power of the Planck CMB constraint is robust to the choice of effective mass definition used in the cosmology model, giving similar results from the thermal and Dodelson-Widrow mechanisms. However, if we allow the lepton asymmetry to be very large ($L=10^{-2}$), the Planck exclusion region is significantly reduced.

We also show the oscillation experiment constraints in the cosmology parameter space, where the same effect is observed. In this parameter space the MINOS constraints rule out a larger fraction of the region allowed by the CMB. 

\section*{Acknowledgements}

We are grateful to Thomas Tram (ICG Portsmouth) for help running the LASAGNA code. We thank Joe Zuntz and Richard Battye (Manchester), and Steen Hannestad (Aarhus) for helpful discussions. This work has been supported by the Science and Technology Facilities Council, the Royal Society, and the European Research Council.





\bibliographystyle{model1-num-names}
\bibliography{planckminos}







\end{document}